\documentclass[prb,aps,nofootinbib, twocolumn, amssymb,superscriptaddress,longbibliography,10pt]{revtex4-2}
\usepackage{amsmath}
\usepackage{amssymb}
\usepackage{amsthm}
\usepackage{amsfonts}%
\usepackage{enumerate}
\usepackage{latexsym}
\usepackage{color}
\usepackage{setspace}
\usepackage{blindtext}
\usepackage{dsfont}
\usepackage{mathrsfs}
\usepackage[normalem]{ulem}

\usepackage{tabularx}
\newcolumntype{Y}{>{\centering\arraybackslash}X}

\usepackage{stackengine}

\usepackage{bm}
\usepackage{graphicx}
\usepackage{subfigure}

\usepackage{hyperref}
\hypersetup{
pdfnewwindow=true, colorlinks=true,
linkcolor=blue, anchorcolor=blue,
citecolor=blue, filecolor=blue,
menucolor=blue, urlcolor=blue}

\usepackage{pifont}

\newcommand{\vk}{\mathbf{k}}

\begin{document}

\title{Topological Kondo semimetals emulated in hetero-bilayer transition metal dichalcogenides}

\author{Fang Xie}
\thanks{These two authors contributed equally.}
\affiliation{Department of Physics \& Astronomy,  Rice Center for Quantum Materials, Rice University, Houston, Texas 77005, USA}
\affiliation{Rice Academy of Fellows, Rice University, Houston, Texas 77005, USA}
\author{Lei Chen}
\thanks{These two authors contributed equally.}
\affiliation{Department of Physics \& Astronomy,  Rice Center for Quantum Materials, Rice University, Houston, Texas 77005, USA}
\author{Yuan Fang}
\affiliation{Department of Physics \& Astronomy,  Rice Center for Quantum Materials, Rice University, Houston, Texas 77005, USA}
\author{Qimiao Si}
\thanks{\href{qmsi@rice.edu}{qmsi@rice.edu}}
\affiliation{Department of Physics \& Astronomy,  Rice Center for Quantum Materials, Rice University, Houston, Texas 77005, USA}

\date{\today}

\begin{abstract}
    The moir\'{e} structure of AB-stacked $\rm{MoTe_2/WSe_2}$ represents a natural platform to realize Kondo lattice models, due to the discrepancy of the bandwidth between the individual layers.
    Here, we study this system at the commensurate filling of $\nu_{\rm tot}=2$.
    Our focus is on the $1+1$ filling
    setting of $\nu_{\rm Mo} =\nu_{\rm W}=1$, which enables a Kondo lattice description.  We find a Kondo semimetal due to the sizable intra-orbital hopping among the electrons in the $\rm{MoTe_2}$ layer.
    The Kondo-driven (emergent) flat band is naturally pinned to the Fermi energy. When combined with the inherent topology of the electronic structure, a topological Kondo semimetal phase ensues.
    We calculate the valley Hall response and, due to the breaking of inversion symmetry, also identify a spontaneous Hall effect. There is a Berry curvature dodecapole, which leads to a fourth-order spontaneous Hall effect in the perturbative regime of the electric field that is further amplified in the non-perturbative regime.
    As such, the system provides a tunable setting to simulate topological Kondo semimetals.
    Finally, we discuss the pathways that connect the physics realized here to the Weyl Kondo semimetals and their proximate phases that have been  advanced in recent years in topological Kondo lattice models and materials.
\end{abstract}

\maketitle

\section{Introduction}\label{sec:intro}

Materials with moir\'e potentials provide an ideal platform to study the interplay between topology and correlation physics, as exemplified by the magic-angle twisted bilayer graphene~\cite{cao_correlated_2018, cao_unconventional_2018}. Moir\'e transition metal dichalcogenides (TMDCs) represent another important example.
Changing the voltage applied to the top and bottom gates enables a continuous tuning of the carrier density and band structure. This tuning process allows for modifications in both the band topology and the interaction strength as measured by the bandwidth, which are among the basic ingredients for realizing novel physics.
A wide variety of emergent phases come into play, including the Chern insulator, fractional Chern insulator, Mott insulator, Wigner crystals, unconventional superconductors, and Kondo effect~\cite{zhao2022gate,Zhao2024np,
Zhao2023.x,dalal_orbitally_2021,Kumar2022Gate,Guerci2023Chiral,Xie2024,Yang2023Metal,Li2021Quantum,Li2021Continuous,ghiotto_quantum_2021,Zeng2023Integer,Park2023Observation, Cai2023Signatures,xu-observation-2023,wu_hubbard_2018,Zhang2021Spin,zang_2021_hartree,zang_dynamical_2022,Devakul2022Quantum,Dong2023Excitonic,morales-duran_nonlocal_2022,Pan2022Topological,Rademaker2022Spin,Crepel2023Topological,Dong2023Composite,yu2023fractional,Xie2022Topological,Xia2024Unconventional,Guo2024Superconductivity}.

\begin{figure*}[t]
    \centering
    \includegraphics[width=\linewidth]{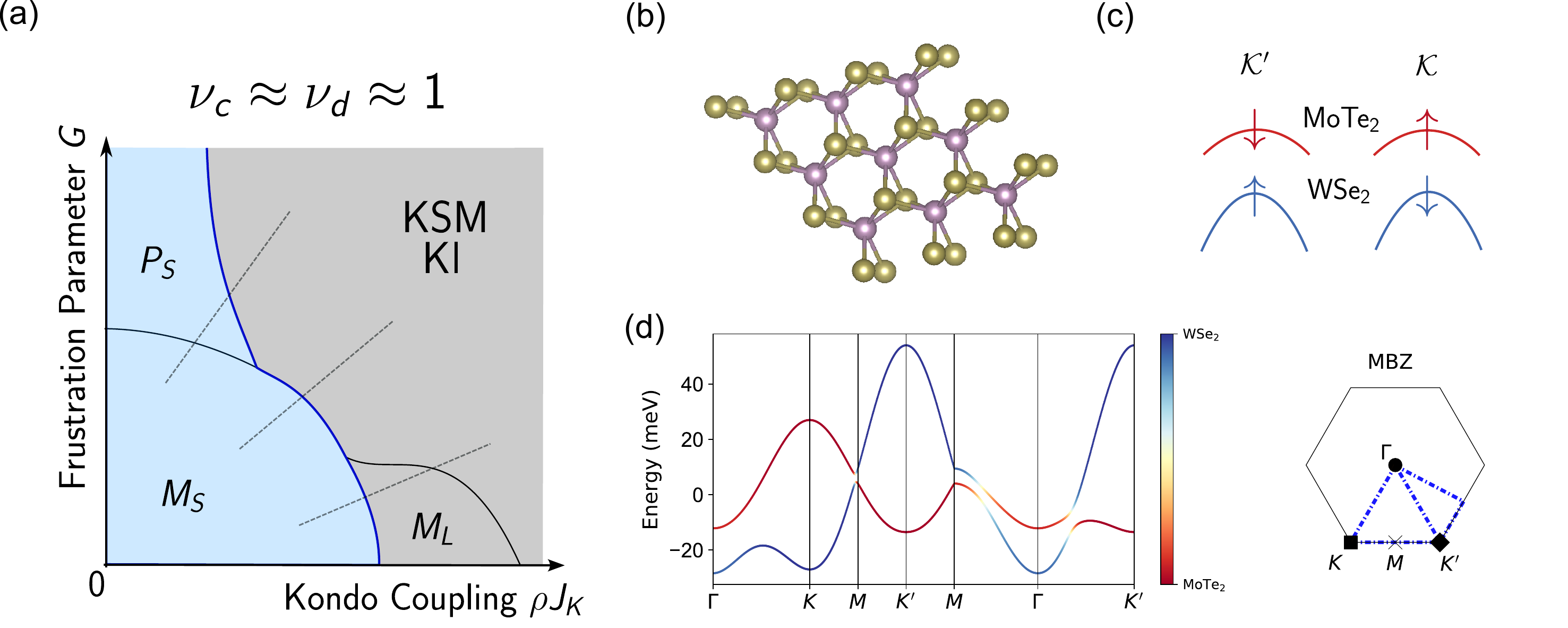}
    \caption{(a) Global phase diagram of Kondo lattice models at commensurate filling factor with ($\nu_c\approx\nu_d\approx1$), where $c$ and $d$ correspond to light and heavy orbitals, respectively.
    Vertical axis $G$ stands for the frustration strength, and horizontal axis $\rho J_K$ describes the Kondo coupling. The blue and black lines correspond to Kondo-destruction (dehybridization) and magnetic transitions, respectively. They separate the parameter region into four phases: (i) $M_S$, a magnetic ordered phase with a small Fermi surface; (ii) $P_S$, a paramagnetic phase with a small Fermi surface; (iii)  $M_L$, a magnetic ordered phase with a large Fermi surface;  and (iv) Kondo-screened phase, which is either a Kondo semimetal (KSM) or a Kondo insulator (KI).
    The dashed lines describe distinct sequences of  quantum phase transitions between the Kondo-destroyed phases and the Kondo-screened ones.
    Adapted from Ref.~\cite{Si13.1}.
    (b) The lattice structure of the $2H$ phase of single layer transition metal dichalcogenides. The purple and yellow orbs represent the metal and chalcogen atoms, respectively.
    This lattice structure already breaks the inversion symmetry.
    Lattice structure data retrieved from the Materials Project for Te2Mo (mp-1023935) from database version v2023.11.1 \cite{MaterialsProject}
    (c) Spin-valley locking in the TMDC materials. $\mathcal{K}, \mathcal{K}'$ denote the two valleys in the TMDC system. Due to the AB-stacking structure, the spin polarizations from the two TMDC layers in the same valley are different, thereby reducing the interlayer hybridization.
    (d) Noninteracting band structure of the two-orbital effective tight binding model with $t_c = 9~\rm meV$, $t_d=4.5~\rm meV$, $t_{cd} = 1.5~\rm meV$ and $\epsilon_{D} = 0~\rm meV$, which well-captures the properties of the active moir\'e bands in AB-stacked $\rm MoTe_2/WSe_2$.
    The definition of the high-symmetry points and high-symmetry lines are also shown.
    }
    \label{fig:intro}
\end{figure*}

Particular interest is focused on a class of moir\'e compounds consisting of AB-stacked heterobilayer TMDCs, where two different kinds of TMDC materials are stacked in an opposite direction~\cite{zhao2022gate,Zhao2024np,Zhao2023.x,Guerci2023Chiral,Xie2024,Yang2023Metal}.
In the present work, we propose this system as an emulator of the topological Kondo semimetals that have in recent years been advanced, both theoretically~\cite{Lai2018,Grefe-prb20.1,Chen-Natphys22} and experimentally~\cite{Dzsaber2017,Dzsaber2021}, with the prospect of realizing a global phase diagram for such commensurately-filled Kondo systems~\cite{Si13.1}.
This phase diagram is illustrated in Fig.~\ref{fig:intro} (a), which captures the interplay between the Kondo effect, magnetism and Kondo destruction that turns the Fermi surface from ``large'' to ``small'' (see below).

\subsection{Electronic structure of AB-stacked \texorpdfstring{$\rm{MoTe_2/WSe_2}$}{MoTe2/WSe2}}

We begin by noting that a single layer of $\rm{MoTe_2}$ or $\rm{WSe_2}$ is in the $2H$ structural phase, which has the space group $P\bar{6}m2$ (No.~187) and time-reversal symmetry.
As shown in Fig.~\ref{fig:intro}(b), each layer has hexagonal unit cells, with metal and chalcogen atoms located at two different sublattice sites, naturally breaking the inversion symmetry.
For the bilayer structure, the space group is reduced to $P3$ (No.~143) which is generated by translations, three-fold rotation $C_{3z}$ and time-reversal symmetry. Its magnetic point group is $31'$, which contains $C_{3z}$ and time-reversal as generators.

The single-layer valence band maxima are typically located at the corners ($\mathcal{K}$ and $\mathcal{K'}$) of the hexagonal Brillouin zone. Moreover, the spin-orbit coupling (SOC) is large, which locks the polarizations of the spin and valley~\cite{Liu2013Three,Kormanyos2015kp} [see Fig.~\ref{fig:intro}(c) for each species]. When different kinds of TMDCs are stacked, the small lattice mismatch between the two components introduces a moir\'e potential that acts on both layers, leading to active moir\'e bands near the Fermi energy.

In the AB-stacked $\rm{MoTe_2/WSe_2}$ platform, over a wide range of parameters, the top two valence bands, which cross the Fermi energy, can be represented by effective Wannier states situated on hexagonal moir\'e superlattice sites~\cite{Zhang2021Spin,Rademaker2022Spin}.
These two Wannier orbitals have distinct kinetic energies because their components come from different TMDC layers. Specifically, the Wannier orbital dominated by the $\rm{MoTe_2}$ layer, which we will denote as $d$ orbital, has a smaller bandwidth than the one dominated by the $\rm{WSe_2}$ layer, which we will mark as $c$ orbital. This contrast can be seen in Fig.~\ref{fig:intro}(d) [see also Fig.~\ref{fig:intro}(a)].

We note on two features that will be particularly important for our considerations below. First, the noninteracting band structure is inherently topological~\cite{Zhang2021Spin, Guerci2023Chiral}. Second, as already noted, the system breaks inversion symmetry.

\subsection{Kondo effect and its realization in heterobilayer TMDCs}

Kondo lattice systems \cite{hewson1997kondo} represent a canonical setting to realize a landscape of correlated quantum phases.
The contrast in the width of the $f$-electron band and that of the $spd$-electron bands causes orbital-selective Mott correlations~\cite{Wirth2016,Kirchner2020}.
In particular, metallic quantum criticality and strange metallicity have been the main stay of the modern era of this field, fueled by the notion of Kondo destruction~\cite{Hu-Natphys2024,Si2001,Colemanetal,senthil2004a}.
This notion captures the dynamical competition between the Kondo and Ruderman–Kittel–Kasuya–Yosida (RKKY) interactions, leading, for example, to an interacting quantum critical point with dynamical ``Planckian" scaling~\cite{Si2001}.
The Kondo destruction notion is supported by a large body of experiments in a variety of quantum critical heavy fermion metals
~\cite{Schroder,paschen2004,Friedemann09,shishido2005,park-nature06,Prochaska2020, Chen2023Shot}.
When combined with geometrical frustration and related quantum fluctuations of the local moments, this notion leads to a global phase diagram~\cite{Si-physicab-06,Si_PSSB10,Coleman_Nev,Pixley-2014}, delineating the interplay between the local-moment magnetism and paramagnetism, the ``large'' Fermi surface associated with the Kondo-screened ground state, and the ``small'' Fermi surface associated with the destruction of the Kondo effect~\cite{paschen_quantum_2021,Hu-qcm2022}.

Previous experimental studies on the AB-stacked heterobilayer $\rm{MoTe_2/WSe_2}$ mainly focused on the $\nu_{\rm Mo} +\nu_{\rm W}=1+x$ region~\cite{zhao2022gate}, where the more localized electrons in the $\rm{MoTe_2}$ are fixed to be at half filling while the carriers in the itinerant $\rm{WSe_2}$ layer are dilute (i.e., $x$ is relatively small).
An analogy with heavy fermion systems has been uncovered in these experiments, as evidenced by a characteristic temperature dependence of the electrical resistivity. At low temperatures, these experiments implied a large effective mass and a suppression of the heavy fermion state by an applied magnetic field.

Suppose we denote the intra-layer hopping between the heavy and light orbitals as $t_d$ and $t_c$, respectively, and the interlayer hybridization as $t_{cd}$.
It has been recognized that, in the $\rm{MoTe_2/WSe_2}$ case, $t_d$ is not that small.
This discrepancy becomes more significant when comparing the hybridization strength $t_{cd}$ between the heavy and light orbitals.
Since interlayer tunneling of the two layers requires breaking spin-$S_z$ conservation, the value of such hybridization is expected to be small~\cite{Zhang2021Spin,Xie2024}.

In other words, if we compare the energy scale in the AB-stacked heterobilayer TMDC system, it will fall into the parameter regime with $t_c > t_d > \tilde{t}_{cd}$ (where $\tilde{t}_{cd} \approx \pi \rho_{0,c}t^2_{cd}$ with $\rho_{0,c}$ being the typical bare density of states of the $c$ band).
This is different from what happens in the canonical heavy fermion systems~\cite{hewson1997kondo,Wirth2016,Kirchner2020} or in the effective Kondo lattice models realized in geometrically-enabled-flat-band systems~\cite{chen2023Metallic,Chen-emergent-2024,HuSciAdv2023}, where $t_c > \tilde{t}_{cd} > t_d$.
Nonetheless, the Coulomb interaction can significantly amplify the renormalized bandwidths difference between the $d$ and $c$ orbitals, with the $d$ orbital approaching the local spin moment limit.
Additionally, the presence of hybridization $\tilde{t}_{cd}$, combined with the fact that the $d$ orbitals form a triangular lattice with geometric frustration, makes this system a potential simulator of Kondo physics.

We also note that the ``inverted" hierarchy of energy scales has important consequences.
Such a large value of $t_d$ obviously leads to strong local charge fluctuations, which means that the Kondo-driven correlated phases are in the so-called ``mixed valence'' regime.
For brevity, it is still usually referred to as Kondo insulators or Kondo semimetals in the literature, as the Kondo effect and Kondo destruction phase transition can still be well-defined in this regime \cite{Si13.1, Checkelsky2024}.
Another concerns the coherent temperature scale, which characterizes the crossover between the high-temperature incoherent scattering regime and the low temperature Fermi liquid regime.
Due to the inverted energy scale hierarchy, the coherence temperature relevant to the strongly correlated metal region depends not only on the hybridization but also on the contribution from the intra-orbital kinetic energy of the heavy electron itself~\cite{Xie2024}.
Accordingly, this coherent temperature scale will be unexpectedly large, as indeed observed experimentally~\cite{zhao2022gate}. As we will show, this energy scale hierarchy plays an important role in realizing a topological Kondo semimetal phase in the commensurate filling that we will study.

\subsection{Topological Kondo lattice Systems at commensurate fillings: Topological Kondo semimetals}\label{sec:kondo-commensurate}

In recent years, topological Kondo lattice models~\cite{Lai2018,Grefe-prb20.1,Chen-Natphys22} and materials~\cite{Dzsaber2017,Dzsaber2021}
have been advanced, which realize the cooperation between the inherent topology of the noninteracting bands and the correlation physics associated with the Kondo effect at commensurate electron fillings.
At such fillings, a Kondo insulator gap would have been opened \cite{Dzero2010Topological, Takimoto2011SmB6}; its prevention by the the lattice symmetry that underlies the band topology gives rise to Weyl Kondo semimetals.

A number of experimental signatures of the Weyl Kondo semimetals have been considered~\cite{Lai2018,Grefe-prb20.1,Chen-Natphys22,Dzsaber2017,Dzsaber2021}.
A $T^3$ specific heat, with an extremely large prefactor,  reflects the linear nodal dispersion with highly reduced velocity. At the same time, a giant spontaneous Hall response captures the effect of large Berry curvatures near the Fermi energy.

\subsection{Scope of the present work}

Recently, the Kondo effect and its destruction have been theoretically discussed in models for the AB-stacked $\rm{MoTe_2/WSe_2}$~\cite{Guerci2023Chiral,Xie2024}.
The inherently topological nature of the electronic bands raises the prospect that this moir\'{e} structure can emulate topological Kondo lattice systems. It is the objective of this work to explore this prospect.

In our previous work \cite{Xie2024}, we focused on incommensurate fillings $\nu_{\rm tot} = 1+x$ of this $\rm {MoTe_2/WSe_2}$ hetero-bilayer system, especially in the regime with rather small $x$.
As we have emphasized in Sec.~\ref{sec:kondo-commensurate}, the commensurate filling case $\nu_{\rm tot} = 1 + 1$ is always considered to be special in the context of heavy fermion systems, which corresponds to potential Kondo topological semimetal or Kondo insulator phases, instead of metallic phases such as heavy Fermi liquid.
In this work, the $U(1)$-slave spin method is used to study how the electronic band structure evolves as the interaction effect gradually increases~\cite{Yu2012U1slave}.
We find a topological Kondo semimetal phase.

We demonstrate a spontaneous Hall effect from the broken inversion symmetry in this topological Kondo semimetal.
Due to the three-fold rotation symmetry and time-reversal symmetry, a Berry curvature dodecapole develops, leading to a fourth-order spontaneous Hall effect in the perturbative regime of the electric field; the effect is expected to be enhanced further in the non-perturbative regime of the electric field.
In addition, we identify a valley Hall effect.

The remainder of the paper is organized as follows.
In Sec.~\ref{sec:model}, we introduce the effective tight-binding and interacting Hamiltonians that describe the low energy physics of this heterobilayer moir\'e system, as well as the method used to solve the interacting Hamiltonian.
We then present our main results from various perspectives in Sec.~\ref{sec:results}.
Finally, we discuss and conclude our findings in Secs.~\ref{sec:discuss} and \ref{sec:summary}.

\section{Model and method}\label{sec:model}
We now turn to the model~\cite{Xie2024} that will be used in our study.

Consider first the noninteracting part of the Hamiltonian.
The continuum Hamiltonian \cite{Zhang2021Spin}, although more faithfully captures the low-energy physics, is not suitable for the study of the correlation physics due to its complexity.
We employ a two-orbital effective tight-binding model \cite{Rademaker2022Spin, Devakul2022Quantum, Guerci2023Chiral, Dong2023Excitonic, Xie2024} to describe the active moir\'e bands in the AB-stacked $\rm{MoTe_2/WSe_2}$ system.
Such effective tight-binding model can be described in terms of the following Hamiltonian:
\begin{align}\label{eq:h0}
    H_0 &= \sum_{\langle\langle ij\rangle\rangle,\tau}t_{c}e^{\tau i\phi^c_{ij}}c^\dagger_{i\tau}c_{j\tau} + \sum_{\langle\langle ij\rangle\rangle, \tau}t_de^{\tau i\phi^d_{ij}}d^\dagger_{i\tau}d_{j\tau}\nonumber\\
    & + \sum_{\langle ij\rangle,\tau} \left(t_{cd}c^\dagger_{i\tau}d_{j\tau} + {\rm h.c.}\right) + \frac{\epsilon_D}{2}\sum_{i\tau}\left(c^\dagger_{i\tau}c_{i\tau} - d^\dagger_{i\tau}d_{i\tau}\right)\,.
\end{align}
As already specified, we use $c$ and $d$ to denote the light and heavy orbitals that are dominated by the $\rm{WSe_2}$ and $\rm{MoTe_2}$ layers, respectively. $\tau=\pm$ is the valley index corresponding to $\mathcal{K}$ and $\mathcal{K'}$. $\epsilon_D$ represents the potential difference between the two layers, which is experimentally tunable by the displacement field.
${\langle\langle ij\rangle\rangle}$ denotes the next-nearest neighbor sites, representing the shortest intralayer hopping, while $\langle ij \rangle$ represents the nearest neighbor sites, which connects electrons in different layers.
The phase factors associated with the intra-layer hoppings are chosen to be $\phi^c_{ij} = 2\pi/3$ and $\phi^d_{ij} = -2\pi/3$, which respects the $C_3$ rotational symmetry.
These hopping phases will position the top band edges of the $\rm MoTe_2/WSe_2$ bands to be located at the $K$ and $K'$ points in the moir\'e Brillouin zone.
The interlayer hoppings occur along the nearest neighbor directions instead of on-site, leading to the opening of a hybridization gap with nontrivial Berry curvature.
To fit the continuum model, we adopt the parameters with $t_c =9\,\rm{meV}$, $t_d =4.5\,\rm{meV}$ and $t_{cd} =1.5\,\rm{meV}$. The dispersion of the tight-binding model with such parameters is shown in Fig.~\ref{fig:intro} (c).

\begin{figure*}[t!]
    \centering
    \includegraphics[width=\linewidth]{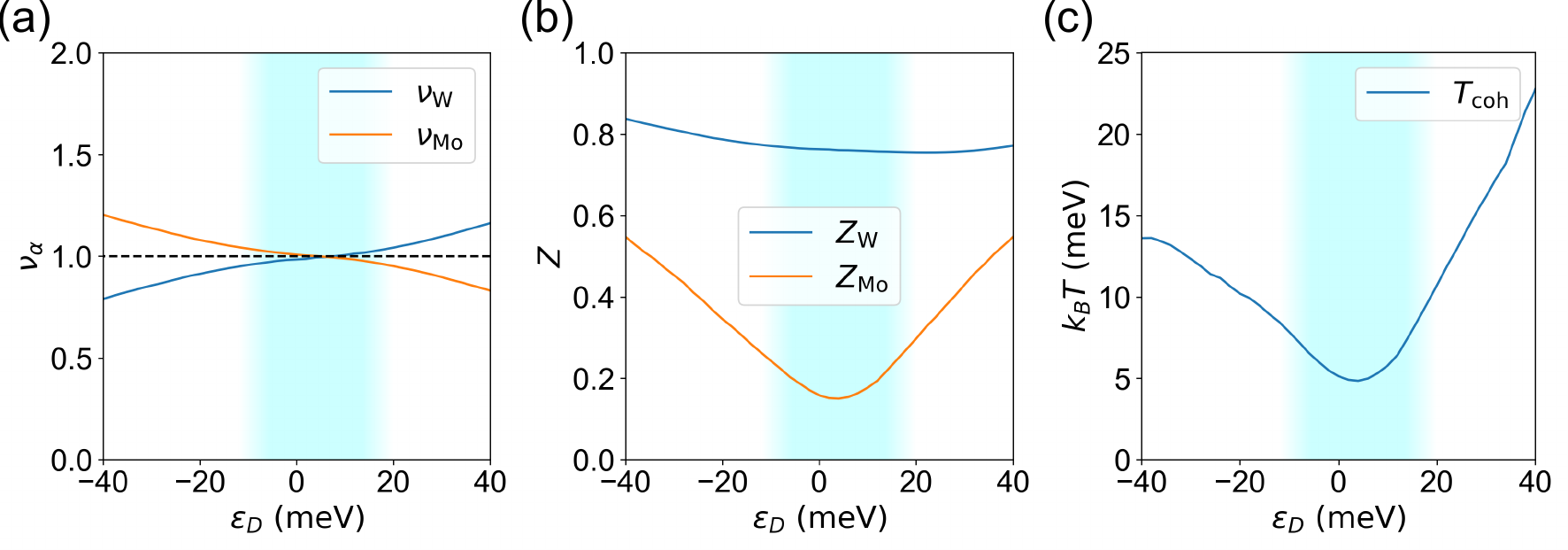}
    \caption{
    (a) The filling factors of the two orbitals
    vs. the displacement field strength.
    (b) The quasiparticle weights of the $d$ and $c$ orbitals,
    for fixed total filling $\nu = 2$.
    (c) The coherence temperature scale.
    In all three panels, the heavy-Fermi liquid region is labeled by the light-blue shadow.
    Here the on-site interaction strength is
   $U = 70\,\rm meV$.
    }
    \label{fig:phase}
\end{figure*}

We next consider the interacting part of the Hamiltonian.
We project the screened Coulomb interaction in terms of the Wannier wave functions obtained from the tight binding model.
Besides the Hubbard (onsite) interaction term, the non-local interactions are not very small compared with the onsite one~\cite{wu_hubbard_2018, morales-duran_nonlocal_2022}.
Therefore, we consider the interactions with a form of extended multiorbital Hubbard-like model as our effective interaction Hamiltonian.
Specifically,
\begin{align}\label{eq:Hint}
    H_I =& \frac{U}{2}\sum_{i,\alpha=c,d}\left(n_{i\alpha+} + n_{i\alpha-} - 1\right)^2 \nonumber\\
    & + V\sum_{\langle ij \rangle, \tau\tau'}\left(n_{ic\tau}-\frac12\right)\left(n_{jd\tau'} - \frac12\right)\nonumber\\
    & +V'\sum_{\langle\langle ij\rangle\rangle,\alpha\tau\tau'} \left(n_{i\alpha\tau} - \frac12\right)\left(n_{j\alpha\tau'} - \frac12\right)\,,
\end{align}
in which $n_{i\alpha \tau} = \alpha^\dagger_{i\tau}\alpha_{i\tau}\,,~\alpha=c{\rm ~or~}d$ is the electron density operator. Here we use $U$ to represent the on-site interaction, $V$ the inter-orbital [nearest-neighbor (NN)] interaction, and $V'$ the next-nearest-neighbor (NNN) interaction. The interaction strength in the lattice model can be estimated from the Wannier functions of the orbitals, and the on-site interaction $U$ is indeed comparable to the bandwidth of the wide band.
In the following study, we will focus on the phase without symmetry breaking, with an emphasis on the orbital-selective correlation effects.

To take into account the interaction effect, we utilize the $U(1)$-slave spin (SS) method~\cite{Yu2012U1slave} to solve the effective Hamiltonian described in Eqs.~(\ref{eq:h0}) and (\ref{eq:Hint}).
We introduce the slave spin representation for a fermion operator as a product of an auxiliary bosonic operator $o^\dagger$ and an auxiliary fermionic operator $f^\dagger$ with $ \alpha^{\dagger}_{i\alpha\tau} = o_{i\alpha\tau}^{\dagger} f^{\dagger}_{i\alpha\tau}$, where $\alpha$ enumerates both $c$ and $d$ electrons.
The auxiliary bosonic field $o^{\dagger}_{i\alpha\tau} = P^{+}_{i\alpha\tau} S^{+}_{i\alpha\tau}P^{-}_{i\alpha\tau}$ is represented by the product of a transverse spin operator $S^+_{i\alpha\tau}$ and projection operators $P^{\pm}_{i\alpha\tau} = \frac{1}{\sqrt{1/2\pm S_{i\alpha\tau}^z}}$, which is adapted to deal with a system away from half filling.
The auxiliary fermion and auxiliary spin are subjected to the constraint that involves the $z$-component of the slave spin operator, $S_{i\alpha\tau}^{z}+\frac{1}{2} = n_{i\alpha\tau}^{f}$, which projects out the unphysical states in the enlarged Hilbert space.
Note that, in the slave spin method, the charge degrees of freedom are represented by the slave spin operators, and the spin (valley here) degrees of freedom are carried by the spinon operators. Based on the aforementioned constraint involving the $S_z$ and $n_{f}$ operators, the density-density type of interactions are exactly represented by the slave spin operators~\cite{Yu2012U1slave}.
In the following, we treat the on-site interaction at the dynamical level using the slave spin operators, and consider the NN and NNN density-density interactions at the mean-field level, as specified by the following equation:
\begin{align}
    H^{S}_{I} =& \sum_i \Bigg{(} \frac{U}{2} \sum_{\alpha} \left( \sum_{\tau} S^z_{i\alpha\tau} \right)^2 + V\sum_{\tau\tau'} ( 3 S_{ic\tau}^z \langle S^z_{d\tau'} \rangle \nonumber\\
    &+ 3 \langle S^{z}_{c\tau}\rangle S^z_{id\tau'}) +V'\sum_{\alpha\tau\tau'}  6 S^{z}_{i\alpha\tau} \langle S^z_{\alpha\tau'}\rangle \Bigg{)}\, ,
\end{align}
where $\langle \cdot \rangle$ denotes the expectation value, which is taken to be translationally invariant.
The quasiparticle weight $Z_{\alpha}$ and the on-site energy renormalization $\Delta\tilde{\varepsilon}_\alpha$ for each orbital $\alpha$ can be solved from such $U(1)$ slave-spin approach.
We refer to the Appendix \ref{app:slave-spin} about the convention and further details of the method as well as on the procedure for numerical solution.

\section{Topological Kondo semimetal}\label{sec:results}

We now present our main results.
We reiterate that our focus is on the strongly correlated setting, at the commensurate filling $\nu_{\rm Mo} + \nu_{\rm W} =1 +1$ [as indicated in Fig.~\ref{fig:phase}(a)].

\subsection{Kondo semimetal}

\begin{figure*}
    \centering
    \includegraphics[width=0.8\linewidth]{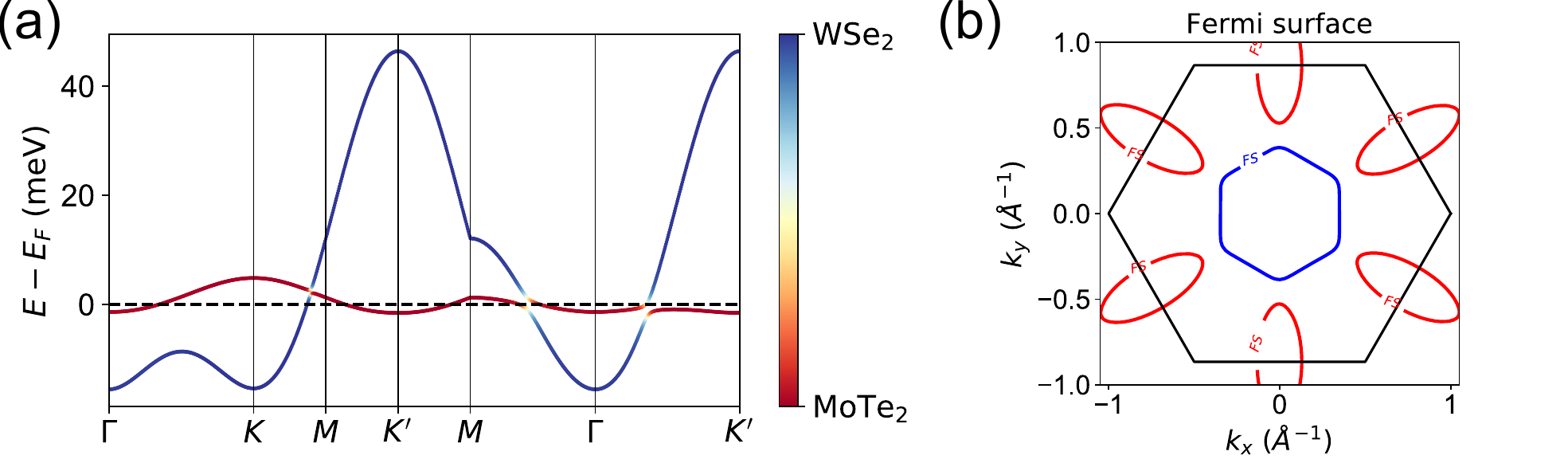}
    \caption{
    (a) The single valley dispersion relation of the renormalized bands along the high symmetry lines of the Brillouin zone in the Kondo semimetal phase.
    The color indicates the ratio of the orbital components.
    (b) Fermi surface contours for $\mathcal{K}$ valley in this Kondo semimetal phase. The blue and red lines stand for electron and hole pockets from the upper and lower bands, respectively. Both (a) and (b) are for the $1+1$ commensurate filling, with the parameters that result in an orbital-selective quasiparticle weight $Z_{\rm{Mo}}=0.16$ and $Z_{\rm{W}}=0.76$; $\epsilon_D=0\,\rm meV$.
    }
    \label{fig:fs}
\end{figure*}

We consider fixed values of the onsite interaction strength $U=70\,\rm{meV}$, NN inter-layer interaction $V = U/2$, and NNN intra-layer interaction $V' = U/4$.
For $\nu_{\rm Mo} \approx 1, \nu_{\rm W} \approx 1$, the potential difference $\epsilon_D$ between the two orbitals is close to zero.

The orbital-resolved filling factors are shown in Fig.~\ref{fig:phase}(a). In the vicinity of $\epsilon_{D}=0$, the filling factor of both orbitals are close to $1$.

The evolution of the quasiparticle weights for both orbitals versus the potential difference $\epsilon_D$ is shown in Fig.~\ref{fig:phase}(b). Clearly, a large quasiparticle renormalization is observed in the $\rm{Mo}$ orbital, in particular in the parameter regime highlighted by the light blue shaded region, where $\epsilon_D$ is close to zero.
The drop in the quasiparticle weight in the heavy orbital is also observed in the correlated metal region when the total filling $1+x$ is close to $1$ (i.e., when $x$ is small)~\cite{Xie2024}.
However, unlike that regime, where the conduction electron is barely renormalized ($Z_{W} \approx 1$), in the $\nu =1+1$ case, the quasiparticle weight number $Z_{W} \approx 0.8$.
This is because the electron filling of the wide $c$ band is close to $1$, making its correlation effect more pronounced.

Fig.~\ref{fig:phase}(c) shows the coherence temperature. It is estimated by taking both the hybridization and  the residual intra-orbital hopping into account~\cite{Xie2024}.
Specifically, the coherence temperature can be estimated through $T_{\rm coh} \sim Z_{\rm Mo}D_{\rm Mo} + \frac{Z_{\rm Mo}Z_{\rm W} t^2_{cd}}{D_{\rm W}}$, in which $D_{\rm Mo} \sim 20 \rm \, meV$ and $D_{\rm W} \sim 40\rm\, meV$ are the half bandwidths of the two active bands.
This value can be much higher than the usual estimation of the coherence temperature from the hybridization effect alone~\cite{nozieres_comments_1998,burdin_coherence_2000,kourris_kondo_2023}.

We now turn to discussing the nature of the phase.
We already mentioned the inverted nature of the underlying energy scales.
Indeed, the width of the noninteracting band for the heavy $d$ ($\rm{MoTe_2}$) orbital rises to about half of that for the light $c$ ($\rm{WSe_2}$) orbital. Consequently, the renormalized dispersion of the heavy orbitals could overcome the hybridization gap, leading to a metallic phase at the commensurate filling.

The renormalized band structure at $\epsilon_D = 0\rm\, meV$ along the high symmetry cut is shown in Fig.~\ref{fig:fs}(a), where one can see that the electron and hole pockets originate from the upper and lower bands, respectively.
As a consequence, the Fermi surface, which is shown in Fig.~\ref{fig:fs}(b), comprises both electron and hole pockets.
We also argue that the qualitative features of the electronic structure is not very sensitive to the value of displacement field strength within this light blue shaded region.
Therefore, in the following, we will mostly use the electronic structure at $\epsilon_D = 0\rm\, meV$ as a representative case.
In Appendix \ref{app:disp-bands}, we present several renormalized band structure plots at different displacement field potential strength.

\subsection{Topological characteristics}

\begin{figure*}[t!]
    \centering
    \includegraphics[width=0.7\linewidth]{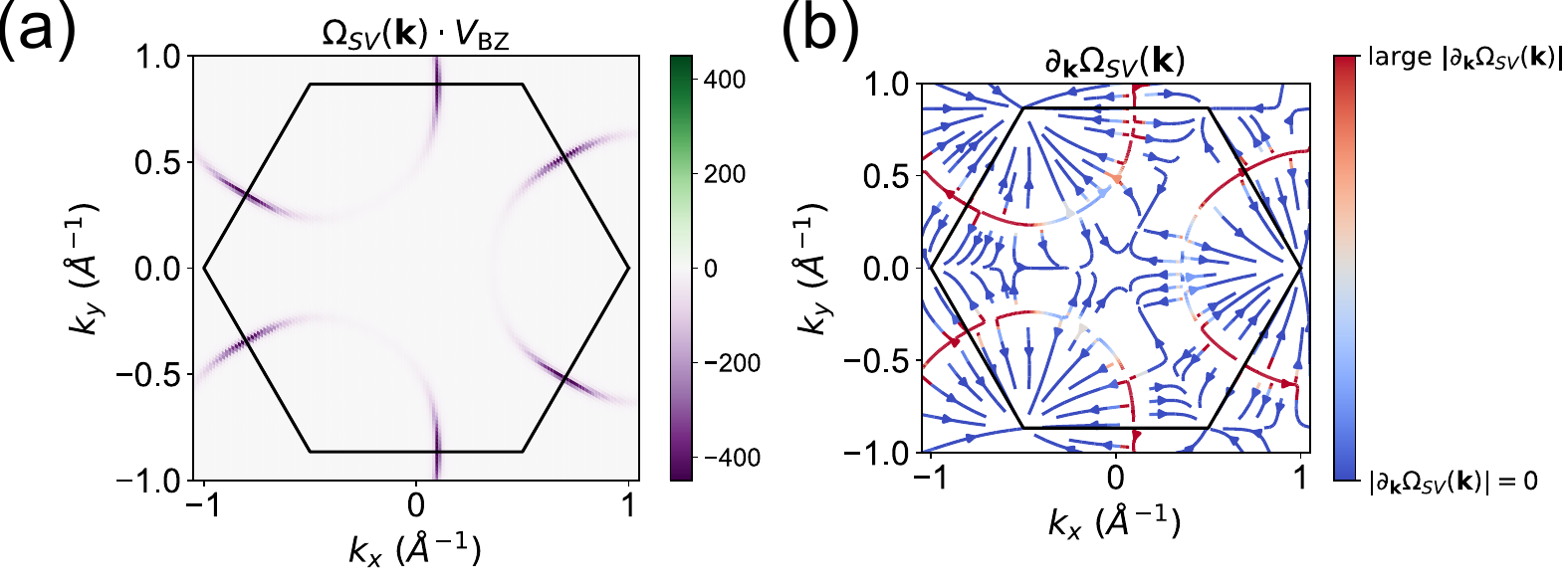}
    \caption{
    (a) Distribution of the Berry curvature of the lower band in a single valley (SV).
    (b) The gradient of the single valley Berry curvature over the first Brillouin zone. The $C_{3z}$ symmetry ensures a vanishing Berry curvature dipole, but the source and sink distribution still indicates a non-vanishing hexapole of the single valley Berry curvature.
    Here we use the same model parameters as in Fig.~\ref{fig:fs}.}
    \label{fig:curvature}
\end{figure*}

To consider the topological characteristics, we start from the single-valley Berry curvature.
The result for the parameters of Fig.~\ref{fig:fs}, with $\epsilon_D = 0$, is presented in Fig.~\ref{fig:curvature}(a).

This single-valley Berry curvature leads to a non-vanishing valley Hall effect.
As seen from Fig.~\ref{fig:fs}(a), while the two bands are fully separated from each other, the indirect gap vanishes (which captures the metallic nature of the system). Accordingly, the valley Hall effect is not quantized.

Based on the Fermi pockets illustrated in Fig.~\ref{fig:fs}(b), we estimated the valley Hall conductivity to be approximately $\sigma_{\rm VH} \approx 0.17 e^2/h$. Although this value may change with model parameter adjustments, it nonetheless indicates the possibility of a sizable valley Hall effect.

A Berry curvature multipole is a tensor defined as $Q_{\alpha_1\dots \alpha_n}=\int_{\vk} f({\epsilon}) \partial_{\alpha_1}\dots \partial_{\alpha_n}\Omega({\vk})$ where $f({\epsilon})$ is the Fermi-Dirac distribution.
The $C_{3z}$ symmetry enforces Berry curvature dipole to vanish, and the non-vanishing Berry curvature hexapole to take the form of $Q_{xx}=Q_{yy}\sim \int_{\vk} \nabla^2_{\vk} \Omega_{SV}$~\cite{Zhang2023Higher}. This allows us to interpret the Berry curvature and Berry curvature hexapole densities as the ``static electric potential'' and the ``charge density''.
The ``electric field'' (Berry curvature dipole density) is shown in Fig.~\ref{fig:curvature}(b).
This non-vanishing Berry curvature hexapole leads to a third order component of the valley Hall effect.

\subsection{Spontaneous Hall Effect}

\begin{figure}[b!]
    \centering
    \includegraphics[width=0.7\linewidth]{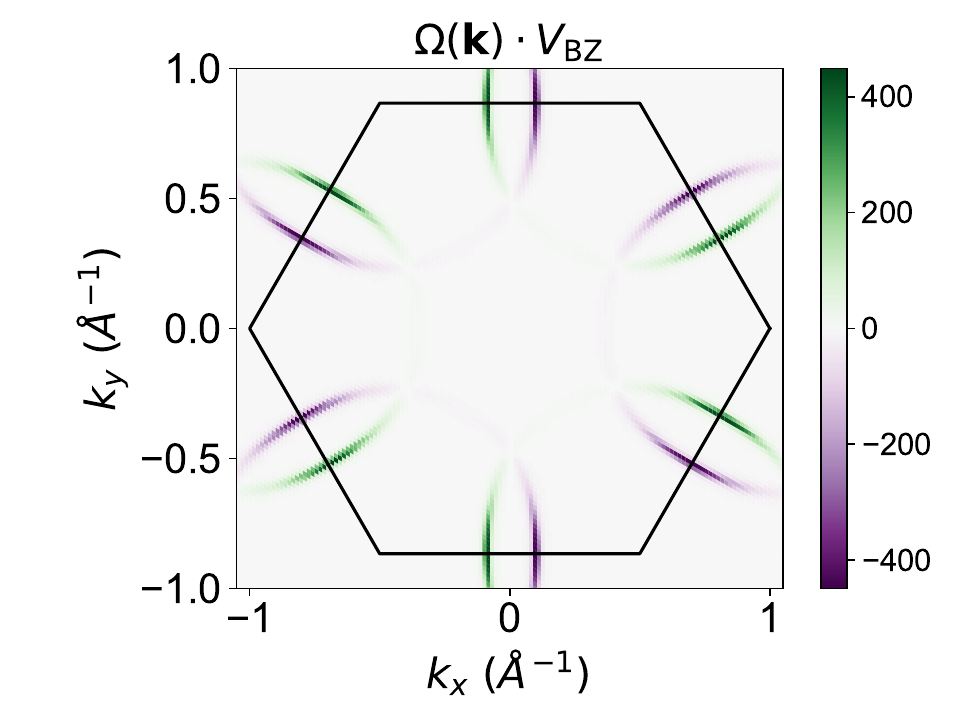}
    \caption{The total Berry curvature with both valleys considered together.
    The model parameters are the same as in Figs.~\ref{fig:fs} and \ref{fig:curvature}.
    }
    \label{fig:total-curvature}
\end{figure}

When both valleys are considered, the Berry curvature distribution is shown in Fig.~\ref{fig:total-curvature}.
A nonzero Berry curvature dipole would lead to a nonlinear Hall response~\cite{Sodemann2015, Deyo2009Semiclassical, Moore2010Confinement}.
More generally, a Berry curvature multipole is responsible for such a spontaneous Hall effect.

The symmetry constraint on a Berry curvature multipole that is covariant under group $G$ is the following condition~\cite{fang2023quantum}:
\begin{equation}
\label{eq:symmetry_constraint}
    Q = \frac{1}{|G|} \sum_{g\in G} g Q\,,
\end{equation}
where $|G|$ is the order of $G$.
Here each component of tensor $Q$ transforms as a vector under $g$.
For point group $G=31'$, Eq.~(\ref{eq:symmetry_constraint}) projects Berry curvature dipole $\int_{\vk} \nabla \Omega$ and Berry curvature hexapole $\int_{\vk} \nabla\nabla \Omega$ to zero; this reflects the $C_{3z}$ and time-reversal symmetry.

Instead, the leading order of Berry curvature multipole is dodecapole, which is a 12-fold polar moment $\int_{\vk} \nabla \nabla \nabla \Omega$, as evidenced by the 12 extrema in the Berry curvature distribution shown in Fig.~\ref{fig:total-curvature}.
This dodecapole leads to fourth order spontaneous Hall effect in the perturbative regime of the electric field~\cite{Zhang2023Higher,fang2023quantum}.
In the AC case, this corresponds to $4\omega$ response, where $\omega$ is the frequency of the applied external electromagnetic field.
In the strongly correlated case, the system is expected to readily be in the fully non-equilibrium regime, which could enhance these responses and can also accordingly produce $1\omega$ responses~\cite{Dzsaber2021,Shouvik2024}.

Thus, we propose measurement of the spontaneous Hall effect as an experimental means of probing this
topological semimetal phase. Our results make it clear that probing both the DC and AC responses will be instructive.
The Berry curvature hexaople can be as large as $Q_{xxxxy} \sim 10^3 a_M^3$, where $a_M$ is the moire lattice scale.
The large value of $a_M$ enhances the prospect of experimentally detecting this nonlinear response.
Because it is comparable to the length scale of the nonlinear response $\ell_E=\hbar/(eE\tau)\sim 10^{-7}\,{\rm m}$ for electric field strength $E_{x} \sim 10\, {\rm V\cdot mm^{-1}}$ and relaxation time $\tau\sim 10^{-12}\,{\rm s}$.
However, the exact value of $Q_{xxxxy}$ is sensitive to whether the Fermi surface overlap with the peak of Berry curvature, and whether it can be measured under current experimental capability depends on the detailed band structure, which is beyond the scope of our model study.

\section{Discussion}\label{sec:discuss}

Several remarks are in order. First, we have shown that the Kondo effect and band topology cooperate to generate a topological Kondo semimetal phase at the commensurate filling $\nu_{\rm Mo}=\nu_{\rm W}=1$.
The topology is captured by the valley Berry curvature, which, in our case, is associated with the development of the flat band from the Kondo effect.
This is analogous to what happens in the Weyl Kondo semimetal~\cite{Lai2018,Grefe-prb20.1,Chen-Natphys22,Dzsaber2017,Dzsaber2021}.

Experimentally, for the AB-stacked heterobilayer $\rm{MoTe_2/WSe_2}$, we note that the case when the total hole concentration $\nu_{\rm tot}=2$ has also been studied, providing evidence for the quantum spin Hall effect~\cite{Zhao2024np}.
However, in the studied parameter region, the value of the displacement field is more likely to correspond to the situation with $\nu_{\rm Mo} +\nu_{\rm W}=2+0$. In this scenario, the electrons mostly reside in one of the layers, which is far away from the Mott insulating region, resulting in a band insulator. Our work applies to the case of a larger displacement field in the region of $\nu_{\rm Mo} +\nu_{\rm W}=1+1$.
We note that two independent recent works \cite{Guerci2024Topological, Mendez-Valderrama2024} have explored related physics in similar models for AB-stacked MoTe$_2$/WSe$_2$, though the spontaneous Hall effect is not considered there.

Second, the analogue between the topological Kondo semimetals we discuss here and the notion of the Weyl Kondo semimetal~\cite{Lai2018,Chen-Natphys22,Dzsaber2017,Dzsaber2021} may be even more direct.
If the TMDC bilayers are stacked into a three-dimensional structure, the Kondo driven flat bands can develop Weyl nodes.
The limitation placed by the separation of the two moir\'{e} bands may be overcome using the notion of emergent flat bands, in which the correlation-induced flat band is pinned to the Fermi energy even though the bare flat band is located away from it~\cite{Chen-emergent-2024}.

In this connection, a related setting is provided by $3d$-electron-based systems on frustrated lattices which, through the notion of compact molecular orbitals, can realize topological Kondo lattice models \cite{chen2023Metallic,Chen-emergent-2024,HuSciAdv2023}. While the existing construction has emphasized two dimensional systems corresponding to the kagome lattice and its variants, the construction is expected to apply to three dimensional systems. In this context, the notion of topological Kondo semimetals has recently been discussed in the case of a clover lattice~\cite{Chen-emergent-2024}.
Thus, we can expect the parallel of our analysis for effective topological Kondo lattice models in the cases of three-dimensional frustrated lattice systems.
More generally, our work also connects with a variety of other models for topological Kondo semimetals~\cite{Feng13,Feng2016,Pixley2017,Cook2017,Chang2018}.
All these further highlight the value of emulating topological Kondo semimetals in the moir\'{e} system as we have considered in the present work.

Third, the more localized $d$ orbitals will experience RKKY interactions. These interactions are expected to cause competition against the tendency towards forming the Kondo-driven flat bands. The expected global phase diagram in the commensurate ($1+1$) electron filling regime has been considered in the canonical Kondo lattice model systems~\cite{Si13.1,Pixley2018,Li_Yu_Si_2019}.
As illustrated in Fig.~\ref{fig:intro}(a), the competing phases involve certain symmetry-breaking order and its symmetry-preserving analogue.
Theoretically, a systematic analysis of this competition could be realized through the extended dynamical mean field theory \cite{Hu2022.edmft}; such a study is left for the future.
Experimentally, it would be instructive to explore the Kondo-destruction phenomena ~\cite{Wirth2016,Kirchner2020,paschen_quantum_2021} in this commensurate filling setting.

Finally, our discussion are based on the assumption that there are no magnetic orders or density wave orders with spontaneous translation or time-reversal symmetry breaking.
Although there is currently no experimental evidence for the existence of such states in the relevant parameter regimes, their presence could indeed affect the potential topological responses.
A detailed study of these possibilities lies beyond the scope of the present study based on parton construction, and would be an interesting direction for future work.

\section{Summary}\label{sec:summary}

We have studied the interplay between the correlation physics and band topology in the AB-stacked $\rm{MoTe_2/WSe_2}$ at the commensurate  $1+1$ filling.
A topological semimetal phase has been identified. We show that this phase displays a Berry curvature dodecapole, leading to a fourth-order spontaneous Hall effect in the perturbative regime of the electric field and a larger effect in the fully non-equilibrium regime of the electric field.
We also find a sizable valley Hall effect.
Our results set the stage to address the interplay between the topological Kondo semimetal and proximate phases in the TMDC moir\'e materials.

\begin{acknowledgments}
We thank Jennifer Cano, Kin Fai Mak, Silke Paschen, Jie Shan and Wenjin Zhao for helpful discussions.
This work has primarily been supported by the
NSF Grant No.\ DMR-2220603
(model construction, F.X. and L.C.), by the
Air Force Office of Scientific Research
under Grant No. FA9550-21-1-0356 (conceptualization and model calculation, F.X., L.C. and Y.F.), and by the Robert A. Welch Foundation Grant No. C-1411 and the Vannevar Bush Faculty Fellowship No. ONR-VB N00014-23-1-2870 (Q.S.).
The majority of the computational calculations have been performed on the Shared University Grid at Rice funded by NSF under Grant No.~EIA-0216467, a partnership between Rice University, Sun Microsystems, and Sigma Solutions, Inc., the Big-Data Private-Cloud Research Cyberinfrastructure MRI-award funded by NSF under Grant No. CNS-1338099, and the Advanced Cyberinfrastructure Coordination Ecosystem: Services \& Support (ACCESS) by NSF under Grant No. DMR170109.
Q.S. acknowledges the hospitality of the Aspen Center for Physics, which is supported by NSF grant No. PHY-2210452.
The data that support the findings of this article are openly available \cite{data_set}.

\end{acknowledgments}

\appendix

\section{\texorpdfstring{$U(1)$}{U(1)} slave spin approach}\label{app:slave-spin}

In this appendix, we briefly review the method of $U(1)$ slave-spin approach.
A physical fermion operator $\alpha_{i\alpha\tau}^\dagger$ ($\alpha$ can be $c$ or $d$) is mapped to a fermionic operator $f^\dagger_{i\alpha\tau}$ and a spin operator $o^\dagger_{i\alpha\tau}$:
\begin{equation}\label{eqn:parton}
    \alpha^\dagger_{i\alpha\tau} \rightarrow f^\dagger_{i\alpha\tau}o^\dagger_{i\alpha\tau}\,,
\end{equation}
in which $i$, $\alpha$ and $\tau$ stand for unit cell, orbital and valley indices.
The slave spin operator has the following form:
\begin{equation}\label{eqn:slave_o}
    o^\dagger_{i\alpha\tau} = P^+_{i\alpha\tau} S^+_{i\alpha\tau}P^-_{i\alpha\tau}\,,~~~P^\pm_{i\alpha\tau} = \frac{1}{\sqrt{\frac{1}{2} \pm S^z_{i\alpha\tau}}}\,,
\end{equation}
where $S^\pm$ and $S^z$ operate on the space of spin-$\frac12$.
We note that such parton construction enlarges the local Hilbert space.
Among the four states spanned by the slave spin and slave fermion, only two of them correspond to the physical states:
\begin{align}
    |n_{i\alpha\tau} = 1\rangle \leftrightarrow& | n^f_{i\alpha\tau}=1 \rangle | S^z_{i\alpha\tau} = \uparrow \rangle\,,\\
    |n_{i\alpha\tau} = 0\rangle \leftrightarrow& | n^f_{i\alpha\tau}=0 \rangle | S^z_{i\alpha\tau} = \downarrow \rangle\,,\\
    \varnothing \leftrightarrow& | n^f_{i\alpha\tau}=0 \rangle | S^z_{i\alpha\tau} = \uparrow \rangle\,,\\
    \varnothing \leftrightarrow& | n^f_{i\alpha\tau}=1 \rangle | S^z_{i\alpha\tau} = \downarrow \rangle\,.
\end{align}
Obviously, the local constraint is given by $\langle S^z_{i\alpha\tau} \rangle + 1/2 = \langle n^f_{i\alpha\tau} \rangle$.
On the saddle-point level, such constraint can be employed by adding an extra Lagrange multiplier term into the parton Hamiltonian:
\begin{equation}
    H_\lambda = \sum_{i\alpha\tau} \lambda_{i\alpha\tau} \left(S^z_{i\alpha\tau} + \frac12 - n^f_{i\alpha\tau} \right)\,,
\end{equation}

Using the parton operators to rewrite the Hamiltonian and and applying a mean-field decoupling between the slave fermions and slave spins, the Hamiltonian can be separated into two distinct parts $H^f$ and $H^S$.
In particular, the slave-spin part becomes local terms with the following form:
\begin{equation}
    H^S = H_I^S + \sum_{\alpha\tau} \lambda_\alpha S^z_{\alpha\tau} + \sum_{\alpha\tau}\left[h_\alpha \frac{S^+_{\alpha\tau}}{\sqrt{n^f_\alpha(1 - n^f_\alpha)}} + {\rm h.c.}\right]\,,
\end{equation}
in which $H_I^S$ is defined in Eq.~(\ref{eq:Hint}) in the main text, which stands for the ``on-site'' part of the interaction written using the slave-spin operators
and $n^f_\alpha = \langle f^\dagger_{i\alpha\tau} f_{i\alpha\tau} \rangle$ is the fermion density expectation value of the orbital $\alpha$.
Here we have assumed that the parameters $\lambda$, $h_\alpha$ and $Z$ are independent on unit cell index $i$ and valley index $\tau$, which represents a non-magnetic state without translation symmetry breaking orders.
The quasiparticle weights are defined as the expectation value of the slave spin operators in the following way:
\begin{equation}
    Z_\alpha = \frac{\langle S^+_{\alpha\tau} \rangle}{\sqrt{n^f_{\alpha}(1 - n^f_\alpha)}}\,,
\end{equation}
and the ``bath'' fields are defined as:
\begin{align}
    h_c =& \frac{1}{N}\sum_{\vk} \epsilon_{cd}(\vk) \sqrt{Z_d} \langle f^\dagger_{\vk c\tau} f_{\vk d\tau} \rangle \nonumber \\
    &+ \frac{1}{N}\sum_{\vk}\epsilon_c(\vk)\sqrt{Z_c} \langle f^\dagger_{\vk c \tau} f_{\vk c\tau} \rangle\,,\\
    h_d =& \frac{1}{N}\sum_{\vk} \epsilon_{cd}(\vk) \sqrt{Z_c} \langle f^\dagger_{\vk d\tau} f_{\vk c\tau} \rangle \nonumber \\
    &+ \frac{1}{N}\sum_{\vk}\epsilon_d(\vk)\sqrt{Z_d} \langle f^\dagger_{\vk d \tau} f_{\vk d\tau} \rangle\,,
\end{align}
in which $\epsilon_{c}(\vk)$, $\epsilon_{d}(\vk)$ and $\epsilon_{cd}(\vk)$ stands for the Fourier transformations of the intralayer and interlayer hopping terms, respectively.
On the fermionic side, the mean-field Hamiltonian has the following quadratic form:
\begin{align}
    H^f =& \sum_{\langle\langle ij\rangle\rangle,\tau}t_{c} Z_c e^{\tau i\phi^c_{ij}}f^\dagger_{ic\tau}f_{jc\tau} + \sum_{\langle\langle ij\rangle\rangle, \tau}t_d Z_d e^{\tau i\phi^d_{ij}}f^\dagger_{id\tau}f_{jd\tau}\nonumber\\
    & + \sum_{\langle ij\rangle,\tau} \left(t_{cd} \sqrt{Z_c Z_d}f^\dagger_{ic\tau}f_{jd\tau} + {\rm h.c.}\right) \nonumber \\
    &+ \sum_{i\tau}f^\dagger_{ic\tau}f_{ic\tau}\left(\frac{\epsilon_D}{2} -\lambda_c + \lambda_c^0 \right)\nonumber\\
    & + \sum_{i\tau}f^\dagger_{id\tau}f_{id\tau}\left(-\frac{\epsilon_D}{2} -\lambda_d + \lambda_d^0 \right)\,,
\end{align}
and the parameters $\lambda^0_\alpha$ are defined as:
\begin{equation}
    \lambda_\alpha^0 = -\sqrt{Z_\alpha}|h_\alpha|\frac{2 n_\alpha^f - 1}{n^f_\alpha(1-n^f_\alpha)}\,.
\end{equation}
These terms guarantee the quasiparticle weights of both orbitals will be reduced to $1$ when the interaction terms are turned off.

We note that the quasiparticle weights $Z_\alpha$, which appear in the fermion Hamiltonian $H^f$ as parameters, can be determined from the ground state of the spin Hamiltonian $H^S$.
Similarly, the bath fields $h_\alpha$, which appear in the spin Hamiltonian $H^S$ as parameters, can be determined from the ground state of the fermion Hamiltonian $H^f$.
Additionally, the Lagrange multipliers $\lambda_\alpha$ are solved from the local constraint equation.
Hence, with all these equations taken into account, the parameters $\lambda_\alpha$, $\lambda_\alpha^0$ and $Z_\alpha$ can be determined self-consistently.

\section{Electronic structure under different displacement field strengths}\label{app:disp-bands}

\begin{figure}[t]
    \centering
    \includegraphics[width=\linewidth]{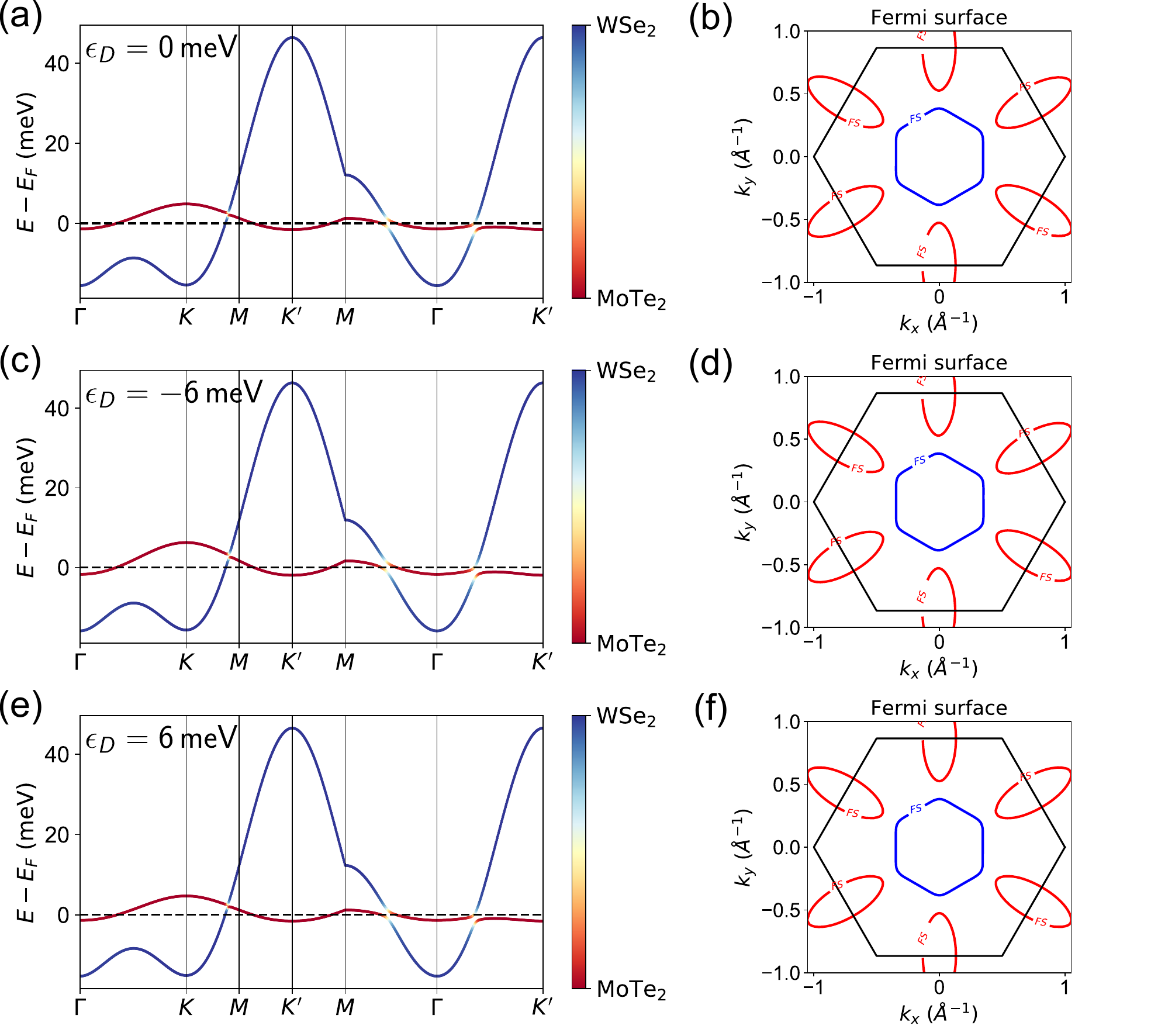}
    \caption{(a-b) The renormalized band structure and the Fermi surface shape in a single valley with displacement field potential strength $\epsilon_D = 0\,\rm meV$.
    (c-d) The renormalized band structure and the Fermi surface in a single valley with displacement field potential strength $\epsilon_D = -6\rm\, meV$.
    (e-f) The single value renormalized band structure and the Fermi surface under displacement field potential strength $\epsilon_D = 6\rm \, meV$.
    Subfigures (a-b) are identical to Fig.~\ref{fig:fs} for easy comparison.
    }
    \label{fig:disp-bands}
\end{figure}

In this appendix, we provide the numerical results of the electronic structure, including the band structure and the Fermi surface structure at the commensurate filling under different displacement field strength.
As shown in Fig.~\ref{fig:phase}, the quasiparticle weight of the Mo orbital is strongly suppressed in the light blue shaded region.
In Figs.~\ref{fig:fs}, \ref{fig:curvature} and \ref{fig:total-curvature}, we show the Berry curvature and its dipole moment at a displacement field strength of $\epsilon_D = 0,\rm meV$, which lies within this interval.

Here we present the renormalized band structure and Fermi surface shape at different displacement field potential strength.
In Figs.~\ref{fig:disp-bands}(c-f), the electronic structure at displacement field potential strength $\epsilon_D = -6\,\rm meV$ and $\epsilon_D = 6\,\rm meV$ are shown.
By comparing these results with the band structure and Fermi surface at $\epsilon_D = 0\,\rm meV$ [shown in Fig.~\ref{fig:disp-bands}(a-b)], we find that the qualitative features, such as the amount of Fermi pockets, remain largely unaffected by the displacement field strength.
However, some quantitative differences can be observed.
For example, due to the slightly different quasiparticle weight of the Mo orbital, its corresponding bandwidth is also slightly narrower at $\epsilon_D = 6\,\rm meV$ than at $\epsilon_D = 0\,\rm meV$.

\bibliography{tmd.bib}

\end{document}